\documentclass[twocolumn,final]{article}
\usepackage{epsfig}
\usepackage{graphicx}
\usepackage{amsmath}
\usepackage{epsfig}
\usepackage{graphics}
\usepackage{graphicx}
\usepackage{amsmath}

\begin{document}

\title{Heat to electricity thermoacoustic-magnetohydrodynamic conversion}
\author{A. A. Castrej\'on-Pita$^\natural$ and G. Huelsz
\\ Centro de Investigaci\'on en Energ\'ia, Universidad Nacional
Aut\'onoma de M\'exico, \\ Ap. P. 34, 62580, Temixco Morelos,
M\'exico}
\date{12/10/2006}

\maketitle
\begin{abstract}
In this work, a new concept for the conversion of heat into
electricity is presented. The conversion is based on the combined
effects of a thermoacoustic prime mover coupled with a
magnetohydrodynamic generator, using different working fluids in
each process. The results of preliminary experiments are also
presented.
\end{abstract}

\section{Introduction}
\footnote{$^\natural$ Present address: Clarendon Laboratory,
University of Oxford, Parks Road, Oxford OX1 3PU, United Kingdom.}
Heat to electricity transducers are devices of enormous importance
in modern society. Most of them use an intermediate step in which
thermal energy is converted into some kind of mechanical energy that
can be used to generate electricity. Since the beginning of the
development of thermoacoustic engines, there have been some
proposals for the use of thermoacoustic prime movers coupled with
magnetohydrodynamic transducers to generate electricity. These
devices can potentially be used as solar energy to electricity
converters and, depending on their efficiency and cost, could be an
alternative to photovoltaic cells. Also, due to its simplicity and
the absence of moving rigid components, thermoacoustic electricity
generation can conveniently be used in space applications.

A thermoacoustic prime mover is a device that converts thermal
energy into mechanical energy in the form of acoustic waves. The
thermoacoustic effect arises due to the thermodynamic interaction
between a compressible fluid and solid surfaces that possesses a
suitable axial temperature gradient. Depending on the form of the
resonator and the location of the temperature gradient, the
thermoacoustic prime mover can generate a standing-wave or a
traveling-wave. An extensive review of the thermoacoustic
principles in the context of engines can be found in
\cite{Swift02}. A magnetohydrodynamic (MHD) transducer uses the
motion of an electrically conducting fluid under an applied
magnetic field to convert the fluid mechanical energy into
electricity. The relative motion of the fluid and the applied
field induces an electric current in the direction transversal to
both the fluid motion and the field, that can be extracted through
electrodes suitably located in the device. MHD electrical
generators have been widely investigated and many interesting
designs have been proposed \cite{Branover93,Branover98}.

Previously, two alternative ways to use thermoacoustic
oscillations to produce alternating electrical current have been
proposed. In the first one, the device is composed by a
thermoacoustic prime mover and a magnetohydrodynamic transducer
(\cite{Swift88} and \cite{Ovando05}). It uses an electrically
conducting liquid as the working fluid for both energy conversion
processes. The disadvantage of this type of heat to electricity
transducer is that high pressures (of the order of 150
atmospheres) for the liquid inside the resonator are needed in
order to generate the acoustic oscillations. In the second
alternative, the oscillatory pressure wave in a gas generated by
the thermoacoustic effect is used to drive an electrically
conducting piston in the presence of a magnetic field to generate
electric power \cite{Backhaus04}.

In this paper, a third alternative for the use of thermoacoustic
oscillations to produce electricity is presented.

\section{Description of the concept}

The alternative proposed in this work is similar to the first one,
it uses a thermoacoustic prime mover and a magnetohydrodynamic
transducer. The originality of this new concept consists in the use
of two working fluids separated by gravity, a gas for the
thermoacoustic prime mover and an electrically conducting liquid for
the magnetohydrodynamic transducer.

The heat to electricity transducer is composed of a thermoacoustic
prime mover, a U-tube partially filled with the liquid and a MHD
Faraday generator. The thermoacoustic prime mover and the liquid
in U-tube form a coupled oscillators system. In the bottom of the
U-tube a conventional MHD Faraday generator is located. The
oscillatory motion of the electrically conducting liquid under the
magnetic field generates an alternate electric potential
difference at the electrodes. In this way, it is possible to
extract electrical power from the generator when a load is
connected to the electrodes through an external circuit.

\section{Preliminary experiments}

\begin{figure}[h]
\begin{center}
\epsfig{width=8.7cm,file=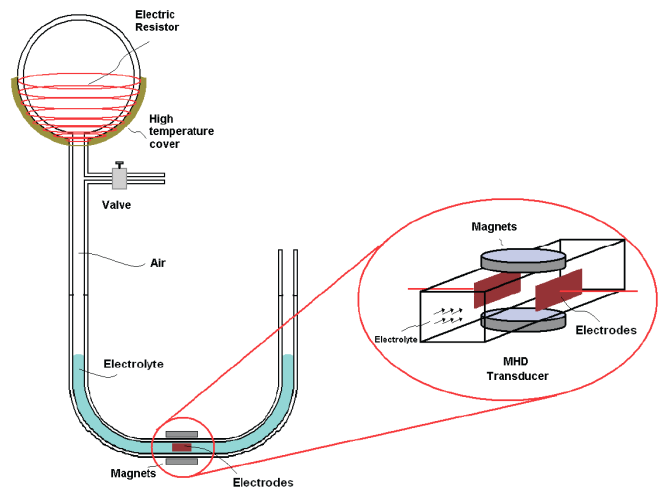} \vspace{-.8cm}
\end{center}
\caption{Sketch of the heat to electricity thermoacoustic-MHD
transducer.} \label{fig1}
\end{figure}

To specify the concept a simple device is constructed. The
thermoacoustic prime mover resonator is a Helmholtz
one\cite{Temkin01}, filled with air. The Helmholtz resonator is
formed by a spherical cavity and a straight tube which is
connected to an U-shaped tube. The lowermost section of the sphere
of the resonator is surrounded by an electrical resistance that
generates a temperature axial gradient on the tube near their
upper part. When the imposed temperature gradient is appropriate,
a thermoacoustic wave is generated in the air inside the Helmholtz
resonator which, in turn, drives an oscillatory motion of an
electrolyte inside the U-tube. In the bottom of the U-tube a
conventional MHD Faraday generator is located. It consists of a
duct of square cross-section where two permanent magnets are
placed in opposite electrically insulated walls in such a way that
the resulting magnetic field is perpendicular to the oscillatory
motion of the electrolyte. Two electrodes are located at the other
pair of walls in the region where the magnetic field is more
intense.  The geometry of the transducer is shown in figure
(\ref{fig1}). The oscillatory motion of the electrolyte under the
magnetic field generates an alternate electric potential
difference at the electrodes. In this way, it is possible to
extract electrical power from the generator when a load is
connected to the electrodes through an external circuit.


The sphere and the tubes used in the experiments (except for the
lower part of the U-tube) are made of soda-lime glass with a wall
thickness of $1.5$ mm. The diameter of the sphere is 38 mm and the
internal diameter of the tubes is 5 mm. The poor thermal
conductivity of the glass helps to generates a large enough
temperature axial gradient in the tube near the heated zone.  The
temperature of the external wall of the heated zone ($T_h$) is
monitored with a thermocouple. The lower part of the U-tube is a
plastic duct of external rectangular cross section of $8.00$ mm $
\times  7.76$ mm and a wall thickness of $0.7$ mm. The total
length of the tubes $L = L_h + L_u + L_o$ is 35 cm, where $L_h$ is
the length of a thermoacoustic tube, $L_u$ is the length of the
liquid column in the U-tube, and $L_o$ is the length of the open
tube filled with air. The electrolyte used for this preliminary
test is a saturated sodium bicarbonate aqueous solution
(electrical conductivity of $6.4 \Omega^{-1} m^{-1}$). Neodyum
magnets of circular cylinder shape, 19 mm in diameter, 10 mm
height, with a maximum magnetic field strength of $0.3$ Tesla at
the center of the circular face, were used. The electrodes are
made of copper plates of $6$ mm long and $4$ mm high placed on
lateral walls inside the tube. The distance between electrodes
($d$) is $6$ mm. Thin wires welded to the electrodes go through
small holes drilled on the plastic duct wall so that an external
circuit can be connected. The potential difference between the two
electrodes was real-time measured with an oscilloscope and
digitalized by an \emph{HP $3709$} data acquisition system. The
axial oscillatory velocity of the electrolyte, at the center of
one of the vertical tubes, was measured using a \emph{DANTEC}
laser Doppler anemometer (LDA). The magnitude of the frequency
spectrum was obtained using Lomb normalized
periodograms~\cite{Press02}.

With a fixed temperature $T_h=350 ^{o}$C, the length of the liquid
column was varied to obtain the maximum liquid column oscillation.
LDA measurements confirmed that maximum oscillation, obtained for a
$L_h = 14$ cm, and $l_u = 11$ cm, was achieved when the oscillation
frequency $5.3 \pm 0.1$ Hz is close to the thermoacoustic optimum
frequency $\sim7$ Hz, given by the diameter of the tube
\cite{Swift02}. The experimental results reported in the following
are obtained for this lengths relation, all this experiments were
made for open circuit conditions.

\begin{figure}[h]
\begin{center}
\epsfig{width=8cm,file=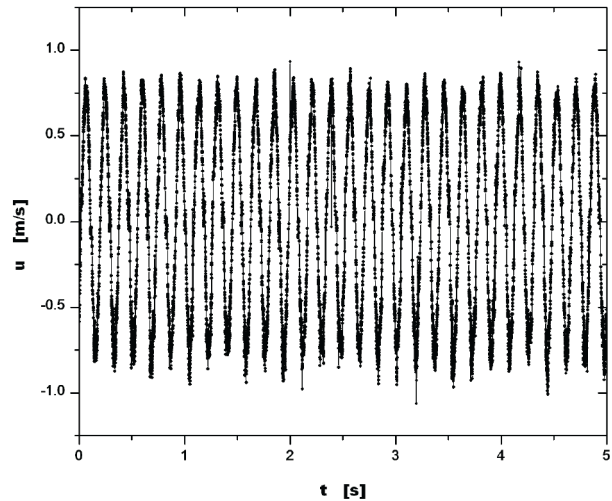} \vspace{-.8cm}
\end{center}
\caption{Axial oscillatory velocity of the electrolyte ($u$), for
 $Th=510  ^{o}$C.} \label{fig2}
\end{figure}

\begin{figure}[h]
\begin{center}
\epsfig{width=8cm,file=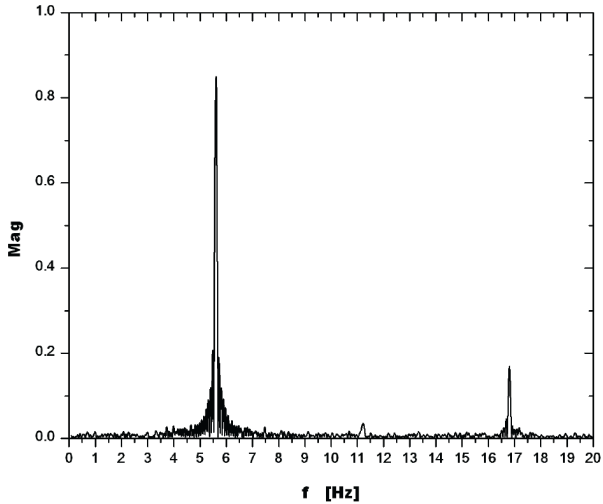} \vspace{-.8cm}
\end{center}
\caption{Frequency spectrum of the axial oscillatory velocity of
the electrolyte ($u$), for  $Th=510  ^{o}$C.} \label{fig3}
\end{figure}

As an example of the LDA measurement results, the axial oscillatory
velocity of the electrolyte ($u$) and its frequency spectrum and
$T_h=510 ^{o}$C, $T_c=30 ^{o}$C conditions are shown in figures
\ref{fig2} and \ref{fig3}. From them, the amplitude of the
oscillatory velocity and the frequency are obtained, being
$u_{a}=0.848 \pm0.005$ m/s and $f_{0}=5.6 \pm 0.1$ Hz respectively,
the second frequency is  $3f_{0}=16.8 \pm 0.1$ Hz and its magnitude
is less than $20\%$ of the magnitude of the fundamental frequency.

\begin{figure}[h]
\begin{center}
\epsfig{width=8cm,file=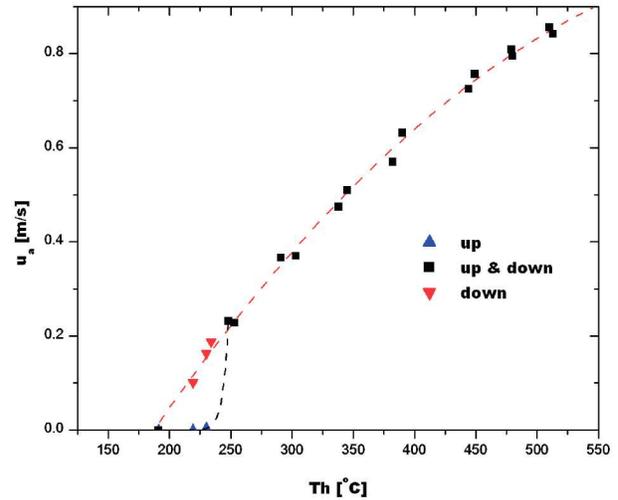} \vspace{-.8cm}
\end{center}
\caption{Amplitude of the axial velocity ($u_{a}$) of the
electrolyte as a function of the temperature $Th$.} \label{fig4}
\end{figure}

The amplitude of the axial velocity ($u_{a}$) of the electrolyte as
a function of the temperature $T_h$ presents hysteresis, as shown in
figure \ref{fig4}. The upper branch is a monotonously increasing
function of the temperature. The frequency of oscillation lightly
depends on $T_h$, from $5.2 \pm 0.1$ Hz for $T_h=180 ^{o}C$ to $5.6
\pm 0.1$ Hz for $T_h=510 ^{o}C$.

\begin{figure}[h]
\begin{center}
\epsfig{width=8cm,file=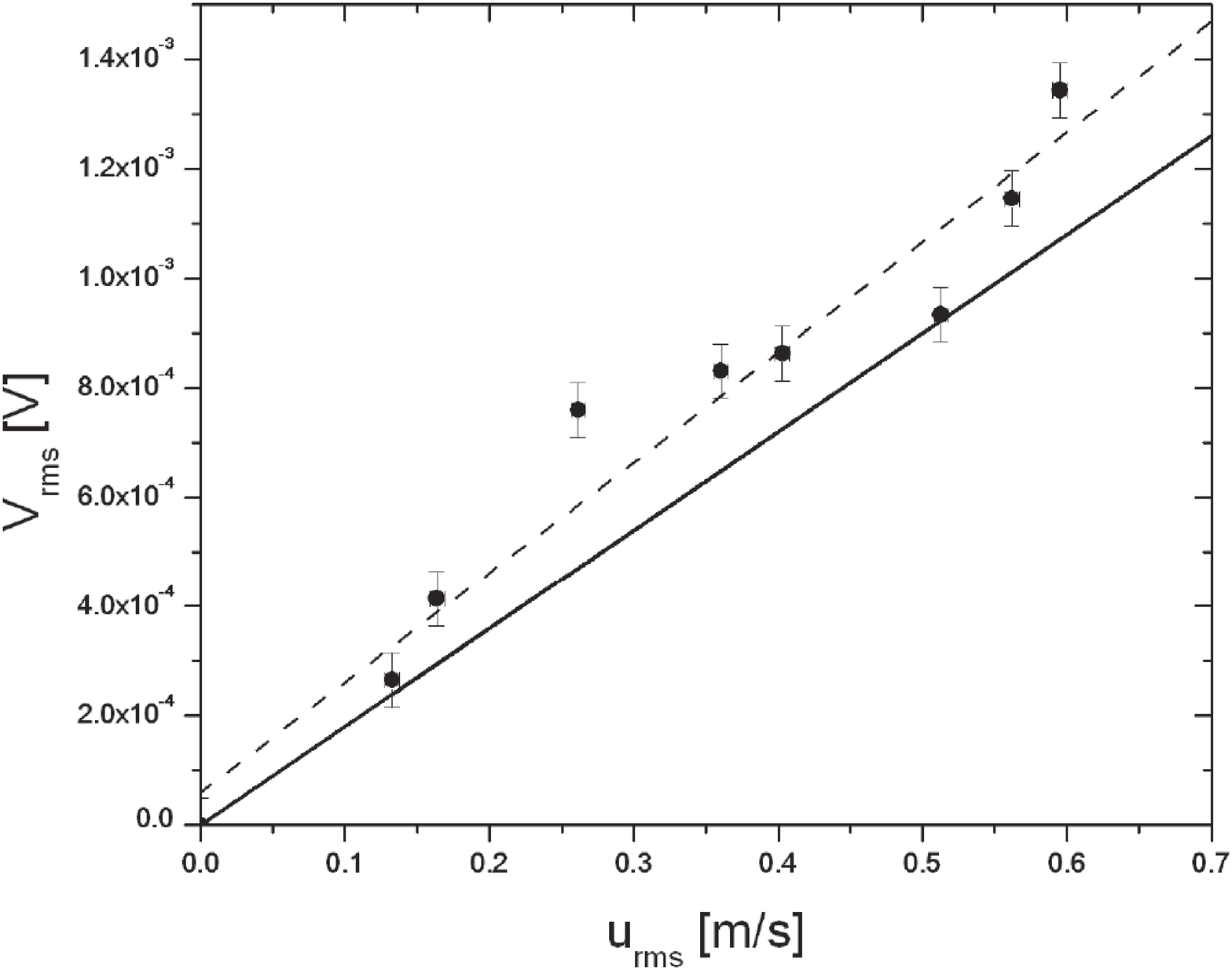} \vspace{-.8cm}
\end{center}
\caption{Root mean square voltage ($V_{rms}$) in open circuit
conditions produced by the device as a function of the root mean
square velocity of the electrolyte ($u_{rms}$). Symbols are
experimental results, the dashed line is the linear fit to them,
the solid line represents theoretical prediction.} \label{fig5}
\end{figure}

In figure \ref{fig5}  the root mean square voltage ($V_{rms}$)
produced by the device as a function of the root mean square
velocity of the electrolyte ($u_{rms}$) is shown; the symbols
correspond to experimental results and the dashed line corresponds
to a linear fit $V_{rms} = 0.00006 + 0.0020 u_{rms}$. The
theoretical curve $V_{rms}=Bdu_{rms}=0.0018 u_{rms}$ is presented
as a solid line \cite{Rosa87}.

It is possible to extract electrical power from the generator when
a load is connected to the electrodes through an external circuit.
Since the voltage generated by this simple device is so small,
measurements of the extracted electrical power as function of the
external load were not performed. A more sophisticated design
using two thermoacoustic prime movers connected through the U-tube
and optimized fluids and geometry is expected to produced a much
higher voltage and consequently, a better overall efficiency.

\section{Conclusions}

A new concept of a heat to electrical power transducer based on the
concurrent action of a thermoacoustically produced compressible wave
and a magnetohydrodynamic transducer, that uses two fluids, a gas
for the thermoacoustic prime mover and an electrical conductor
liquid for the magnetohydrodynamic transducer is presented. In
principle, this device can be used to generate electricity from a
heat source as long as the system is in the presence of gravity or a
centrifugal force. The concept is demonstrated with a simple device,
but a more sophisticated design using two thermoacoustic prime
movers connected through the U-tube and optimized fluids and
geometry, is expected to have a much higher overall efficiency.
Specially, if the natural frequency of the thermoacoustic prime
movers and of the liquid column are commensurable, ideally the same,
the oscillation frequency of the coupled system would be optimized.
This optimal coupling would lead to larger oscillation amplitudes,
and therefore, the efficiency of the device would increase.

\section{ACKNOWLEDGEMENTS}
This work was supported by the CONACyT U41347-F project. The authors
wish to thank Eduardo Ramos and Sergio Cuevas for their comments,
and Margarita Miranda and Guillermo Hern\'{a}ndez for their
technical support.

\end{document}